\documentclass[twocolumn,showpacs,preprintnumbers,amsmath,amssymb]{revtex4}
\usepackage{epsfig}
\preprint{LSCO/2004}
\begin{document}
\title{Frequency-dependent Thermal Response of the Charge System and Restricted Sum Rules in La$_{2-x}$Sr$_{x}$CuO$_4$}
\author{M. Ortolani, P. Calvani, and S. Lupi}
\affiliation{"Coherentia" - INFM and Dipartimento di Fisica, Universit\`a di Roma La Sapienza, Piazzale Aldo Moro 2, I-00185 Roma, Italy}
\date{\today}

\begin{abstract}
By using new and previous measurements of the $ab$-plane conductivity $\sigma_1^{ab} (\omega,T)$ of La$_{2-x}$Sr$_x$CuO$_{4}$ (LSCO) it is shown that the spectral weight $W  = \int_0^\Omega {\sigma_1^{ab} (\omega,T) d\omega}$ obeys the same law $W = W_0 - B(\Omega) T^2$ which holds for a conventional metal like gold, for $\Omega$'s below the plasma frequency. However $B(\Omega)$, which measures the "thermal response" of the charge system, in LSCO exhibits a peculiar behavior which points towards correlation effects. In terms of hopping models, $B(\Omega)$ is directly related to an energy scale $t_T$, smaller by one order of magnitude than the full bandwidth $t_0 \sim W_0$. 
\end{abstract}
\pacs{74.25.Gz, 74.72.-h, 74.25.Kc}
\maketitle

The optical properties of high-$T_c$ cuprates, both in the normal and in the superconducting phase, are still extensively discussed in the literature. For $T > T_c$, experiments where the infrared conductivity $\sigma_1^{ab} (\omega)$ is peaked at $\omega$ = 0 and smoothly decreases with $\omega$ support one-component approaches, like the extended Drude model.\cite{Timusk}  In other experiments, contributions peaked at finite frequencies show up, like polaron-like bands in the mid infrared (MIR)\cite{Lupi99} and resonances in the far-infrared (FIR)\cite{Lucarelli} which point towards charge localization and ordering. In the superconducting phase, contradictory results are reported on the London penetration depth\cite{Tanner93,Tajima_pre} extracted from $\sigma_1^{ab} (\omega)$.

Therefore, recent works have been often focused on a model-independent quantity like the spectral weight

\begin{equation}
W (\Omega,T) = \int_0^\Omega {\sigma_1(\omega,T)d \omega} 
\label{weight}
\end{equation}

\noindent
and on its behavior in different spectral ranges. For $\Omega \to \infty$, the sum rule on the real part of the optical conductivity $\sigma_1(\omega,T)$ requires that  $W (\Omega,T)$ is independent of temperature. However, there are a few special cases where "restricted sum rules" can be considered. In metals $W(\omega_p)$, where $\omega_p$ is the plasma frequency which approximately coincides with the minimum in the reflectivity (plasma edge), is expected to be nearly independent of $T$.\cite{Benfatto_pre} Another example is the Ferrel-Glover-Tinkham (FGT) sum rule which predicts that the $W$ lost at low energy when a superconductor is cooled below $T_c$, is recovered in the $\delta$ function at $\omega$ = 0. 

In cuprates like Bi$_2$Sr$_2$CaCu$_2$O$_{8+y}$ (BSCCO) and YBa$_2$Cu$_3$O$_{7-y}$ (YBCO) the restricted sum rules have been investigated by several groups.\cite{Santander02,Molegraaf,Homes,Boris} In La$_{2-x}$Sr$_x$CuO$_{4}$ (LSCO), a similar study on the in-plane $\sigma_1^{ab} (\omega)$ has recently appeared,\cite{Tajima_pre} while a previous work was focused on the $c$-axis conductivity.\cite{Basov}
In the present paper we study the behavior of $W(\Omega,T)$ in LSCO, both in the normal phase, by comparing the behavior of an underdoped superconductor with $x$ = 0.12 with that of a non-superconducting metal with $x$ = 0.26 and, below $T_c$, in three LSCO samples with different doping. In the former case, starting from the sum rule restricted to $\omega_p$, we will observe topic differences between LSCO, BSCCO and conventional metals, which lead us to identify two different energy scales for the charge system in the cuprates. In the superconducting phase we will find that the FGT sum rule holds if the frequency range is extended to about 4000 cm$^{-1}$.

\begin{figure}
{\hbox{\psfig{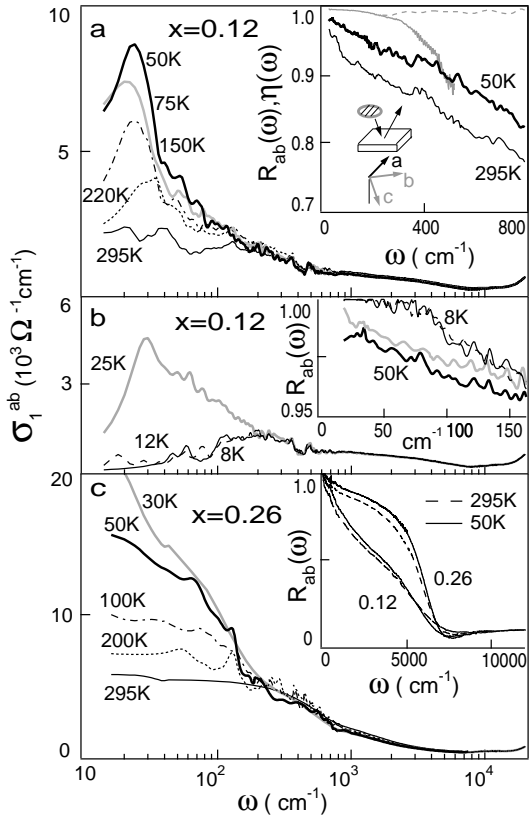}}}
\caption{Optical conductivity, from the reflectivity data reported in the inset, of the underdoped crystal $x$ = 0.12 above $T_c$ (a) and below $T_c$ (b) compared with that of the non-superconducting metal $x$ = 0.26 (c). The insets show the corresponding $R_{ab}$: in the FIR for $x$ = 0.12 (a) above  and (b) below $T_c$; in the whole energy range for both $x$ = 012 and 0.26 (c). In the inset of (a) the grey lines plot the efficiency $\eta$ of the polarizers employed in the experiment: polyethylene (solid) and KRS-5 (dotted). } 
\label{sigma}
\end{figure}

The reflectivity $R_{ab}(\omega)$ of the two LSCO single crystals with $x$ =0.12 and a non-superconducting metal with $x$ = 0.26, was measured with respect to that of a gold film (between 15 and 12000 cm$^{-1}$ ) and a silver film (up to 20000 cm$^{-1}$), both of which evaporated \textit{in situ} onto the sample. The reflectivity of gold was then measured with respect to a platinum mirror, either in order to correct for its $\omega$ dependence in the near infrared, and to compare the response of LSCO at $x$=0.26, which is often indicated as a "normal metal" with that of a good conventional metal.  The optical configuration for measuring $R_{ab}(\omega)$ in LSCO is shown in the inset of Fig.\ \ref{sigma}-a. The radiation impinges under an incidence angle of 8$^0$ on the surface of the crystal. By use of a 4-circle diffractometer and a laser we determined for $x$ =0.12 a miscut of $\theta$ = 1.0$^0$ $\pm$ 0.5$^0$ with respect to an ideal $ab$ plane. Therefore, we used a polarizer to align the electric field orthogonally to the miscut plane. We show elsewhere\cite{unpub} that, under these conditions, the relative deviation with respect to the ideal reflectivity of the $ab$ plane is 

\begin{equation}
\frac {\Delta R_{ab}(\omega)}{R_{ab}(\omega)}\simeq  \sqrt{2(1-\eta)} \frac{1-R^2_{ab}(\omega)}{ R_{ab}(\omega)} \theta^2 g(\omega) \, . 
\label{error-R }
\end{equation}

\noindent
Here, $\eta = (I_{p}-I_{u})/(I_{p}+I_{u})$, as $I_{p}$ ($I_{u}$) is the intensity of the field component parallel (orthogonal) to the polarizing direction. It is plotted in the inset (grey lines) for both the polyethylene and the KRS-5 device here used. Moreover,

\begin{equation}
g(\omega) = \left[\frac{-2 \omega}{\pi \epsilon_2^{ab}} \large \wp \int_0^\infty {\frac {-(\epsilon_1^{ab})^2/2\epsilon_1^{c}}{\omega'^2 - \omega^2} d\omega'}  + \frac{ 1}{2} \right]
\label{g}
\end{equation}

\noindent
$\Delta R_{ab}(\omega)/R_{ab}(\omega)$ turns out to be\cite{unpub} on the order of 0.001 at 100 cm$^{-1}$, much smaller than the experimental noise. In fact, no signature of the $c$-axis phonons appears in the $R_{ab}(\omega)$ of the $x$ = 0.12 sample (see the top inset of Fig.\ \ref{sigma}). In previous measurements on the $ac$ surface of the same sample, reported in Ref. \onlinecite{Lucarelli}, a dip from such a phonon appeared around 470 cm$^{-1}$.\cite{comment} In that case, however, the commercial polyethylene polarizer was used in its whole transmittance range. As shown in the inset, $\eta$ drops to 0.93 at that frequency, causing a 4\% deviation from the real reflectivity of the $a$ axis. Below 200 cm$^{-1}$, where $\eta >$ 0.99, the results of Ref. \onlinecite{Lucarelli} on this and other samples were not affected by the $c$-axis response. This is confirmed by the results with the present safe procedure, that was applied to both $x$ = 0.12 and 0.26. In the former one the in-plane conductivity $\sigma_1^{ab} (\omega)$, once extracted from the $R_{ab}(\omega)$ extrapolated to $\omega = 0$ by a Drude-Lorentz fit, shows again (Fig.\ \ref{sigma}-a) the $T$-dependent resonance at $\simeq$ 30 cm$^{-1}$ reported in Ref. \onlinecite{Lucarelli} for this and other LSCO samples. The resonance is not observed in the non-superconducting metal with $x$ = 0.26 (Fig.\ \ref{sigma}-c). A discussion of the FIR peak and of other details of the LSCO conductivity is reported in Refs. \onlinecite{Lucarelli,Benfatto}.  They are not relevant to the present study of the spectral weight in LSCO, a quantity insensitive to narrow spectral features in the far infrared. 

\begin{figure}
{\hbox{\psfig{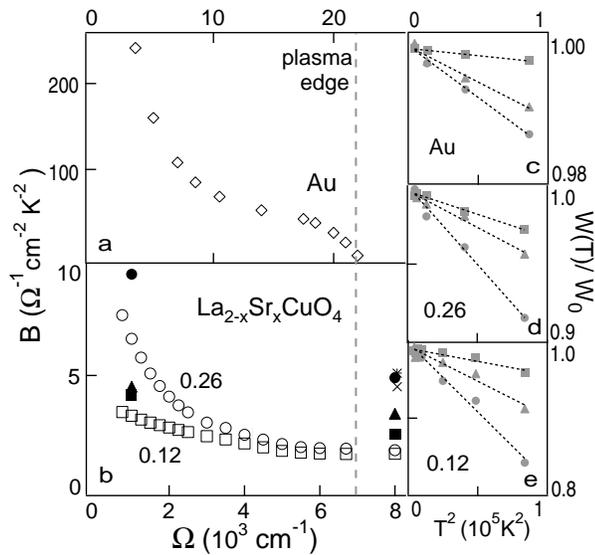}}}
\caption{The coefficient $B (\Omega)$ in Eq.\ \ref{T_square} is plotted for gold (a), and LSCO with $x$ = 0.12 and 0.26 (b), as extracted from plots like those in (c), (d), and (e), respectively. Therein, grey squares refer to $\Omega = \omega_p$, triangles to $\Omega = 0.5 \omega_p$, circles to $\Omega = 0.2 \omega_p$. In (b), $B$ values obtained from the existing data on Bi$_2$Sr$_2$CaCu$_2$O$_{8+y}$ with different $y$ and $T_c$ are reported for comparison: full symbols are from Ref. \onlinecite{Santander02} (circles, $T_c$ = 70 K; triangles, 80 K, squares, 63 K); the star ($T_c$ = 88 K) and the cross (66 K) are extracted from the data of Ref. \onlinecite{Molegraaf}.} 
\label{B}
\end{figure}

In both samples of Fig.\ \ref{sigma}, the plasma frequency obtained from the condition $\epsilon_1 (\omega_p)$ = 0 is 6100 $\pm$ 150 cm$^{-1}$, independent of $T$ within errors. This value is also in agreement with that (6240 cm$^{-1}$) reported for thin LSCO films with different doping.\cite{Tanner93} In conventional metals $W$ exhibits a temperature dependence 

\begin{equation}
W (\omega_p,T) \simeq  W_0 - BT^2 \, ,
\label{T_square}
\end{equation}

\noindent
where $W_0$ accounts for all the carriers in the conduction band, while $B$ depends crucially on the density of states at the Fermi energy $\rho(E_F)$. In a tight-binding approach \textit{both $W_0$ and $B$ depend on the same hopping rate $t_0$}. Eq.\ \ref{T_square} is verified in gold, as one can see from our data in Fig.\ \ref{B}-c where $\omega_p$ = 20500 cm$^{-1}$ (from the zero-crossing of $\epsilon_1$). 

It has been found that Eq.\ \ref{T_square} holds also for a high-$T_c$ superconductor like BSCCO.\cite{Molegraaf,Santander03}  Here, Figs.\ \ref{B}-d and -e show that this behavior is verified also in LSCO, both for $x$ = 0.26 and $x$ = 0.12. One may ask if this findings are sufficient to extend the above tight-binding approach, characterized by a single energy scale $t_0$, to cuprates. To obtain deeper insight, we notice that for LSCO, basing on the good fits in Figs.\ \ref{B}-d and -e, the above $T^2$ dependence is verified not only at $\omega_p$ but also for lower values of $\Omega$ (provided that they are higher than the highest phonon frequency, $\sim$ 700 cm$^{-1}$). Therefore we can write for both gold and LSCO

\begin{equation}
W (\Omega,T) \simeq  W_0 - B(\Omega)T^2 \, .
\label{B_Omega}
\end{equation}

The frequency-dependent coefficient $B(\Omega)$, that we introduce through Eq.\ \ref{B_Omega}, describes the "thermal response" of the carriers. It can be evaluated at any $\Omega$ as done in Figs.\ \ref{B}-c, -d, or -e. The resulting values are reported in the left panels of the same Figure. In gold, (a) all $T$-dependent mechanisms are confined at $\omega \alt \omega_p$ and at the plasma edge, $B \simeq 0$. In both LSCO samples, on the contrary, at the edge $B$ is still much different from zero [$B(\omega_p)$ = 1.7 $\Omega^{-1}$ cm$^{-2}$ K$^{-2}$]. (When evaluating these figures one should consider that the $B$ scale for gold is larger by more than a factor of 10 than in LSCO, as $W_0 \propto \omega_p^2$). The result at $\omega_p$ is not surprising, in view of the correlation effects that, in LSCO, may extend the $T$-dependence of the carrier response up to energies on the order of the Hubbard splitting $U$. One should then observe similar effects in other cuprates. Indeed, by using the data of Ref. \onlinecite{Molegraaf,Santander03} on five BSCCO samples with different $T_c$'s, one obtains the $B$ values shown for comparison in Fig. \ \ref{B}-b. All of them, at the BSCCO $\omega_p \sim$ 8000 cm$^{-1}$, are even higher than here found in LSCO.

One may ask if the predictions of the non-interacting, tight-binding model for the Cu-O square lattice are compatible with these results. For the present purpose one can use a simplified one-band picture and obtain

\begin{equation}
W (\omega_p,T) = \frac {\pi e^2 a^2}{2 h^2 V} K 
\label{K}
\end{equation}

\noindent 
where $K$ is the kinetic energy of the carriers, $a$ the Cu-O plane lattice parameter, and $V$ the LSCO cell volume. In this framework, Eq.\ \ref{T_square} derives directly from the Sommerfeld expansion of the Fermi distribution function\cite{Ashcroft} which, in a first approximation, also gives

\begin{equation}
B (\omega_p) \simeq \frac {\pi e^2 a^2}{2 \hbar^2 V} \frac {\pi^2 k_B^2}{6} \rho(E_F)
\label{rho}
\end{equation}

\noindent
In two dimensions one may introduce the simplifying assumption of a rectangular density of states, so as $\rho(E_F) = \rho(E) = (4 t_T)^{-1}$  where $t_T$ is the hopping rate. Then one finds $t_T = \pi^2/[24B(\omega_p)]$ = 22 meV which corresponds to a bandwidth $8t_T$ = 176 meV. By applying the same procedure, even smaller values for $t_T$ can be extracted from the data\cite{Molegraaf,Santander03} reported in Fig.\ \ref{B} for Bi$_2$Sr$_2$CaCu$_2$O$_{8+y}$.
On the other hand, applying to the present data Eq.\ \ref{K}, one obtains $W_0 \simeq$ 240 meV for $x$ = 0.12 and 500 meV for $x$ = 0.26. This provides the hopping rate related to the full bandwidth, that we call $t_0$. As in any hopping model $t_0 \alt W_0 \alt 2t_0$, $t_0$ is in qualitative agreement both with estimates for the Cu-O planes from photoemission data (250-300 meV),\cite{Norman}and with energy band calculations (430 meV)\cite{Andersen}. However $t_0$ is larger, by one order of magnitude, than the above $t_T$ obtained from the temperature dependence of $W$. In the non-interacting description, the two values should be the same. The present inconsistency shows once again that a simple hopping model cannot describe the electrodynamics of the Cu-O planes. One could perform more reliable tight-binding calculations, including the effects of next-nearest neighbor hopping rate $t'$, to obtain a different value of $\rho(E_F)$ and hence of $t_T$. However it is unlikely that such corrections may increase $t_T$ by an order of magnitude. On the other hand, the possibility that the Fermi level is close to the Van Hove singularity in $\rho (E)$ can be excluded. It would imply that $t_T$ might change very strongly when passing from $x$ = 0.12 to 0.26, in contradiction with the present results. 

In summary, our results indicate a coexistence of \textit{two different energy scales} in Eq.\ \ref{B_Omega}, $t_0$ and $t_T$. The former one is related to the width of the broad conduction band built up either directly by doping and by doping-induced transfer of spectral weight from the high-energy bands.\cite{Uchida} In turn, $t_T$ seems to control the transfer of spectral weight that is triggered by temperature. From this point of view, one may notice that a similar energy scale is involved in pseudogap formation.\cite{Timusk} However, as the latter phenomenon is restricted to underdoped compounds,
that hint should be supported by analyzing the difference in the low-energy thermal response between 0.12 and 0.26, which clearly appears in Fig.\ \ref{B}-b. 

The behavior of $B (\Omega)$ for $x$ = 0.12 in Fig.\ \ref{B} deserves a few further comments. As the $T$-dependence of the response is concentrated below $\sim$ 4000 cm$^{-1}$, the Drude term vanishes at frequencies definitely lower than the pseudo-plasma edge at $\sim$ 6000 cm$^{-1}$. This may explain why this energy is basically insensitive to the number of carriers (in the inset of Fig.\ \ref{sigma}-c it is the same for $x$ = 0.12 and 0.26). 
What is left in the optical conductivity might be the mid-infrared band peaked at $\sim$ 0.5 eV, which can be directly observed in the semiconducting phase of several cuprates\cite{Uchida}  and is usually included in the multi-component models of $\sigma_1^{ab} (\omega)$ as a $T$-independent component.\cite{Lupi2004}  Therefore, in Eq.\ \ref{B_Omega}, it would not contribute to $B(\Omega)$ but might account for a large part of the $T$-independent spectral weight $W_0$.

\begin{figure}
{\hbox{\psfig{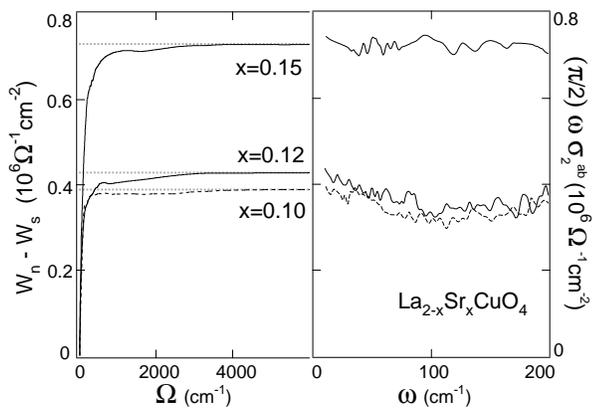}}}
\caption{In the left panel, the difference $W_n - W_s$ between the spectral weight calculated from $\sigma_1^{ab} (\omega)$ at  50 K and that at $T < T_c$ is plotted vs. the integration limit $\Omega$ for $x$ = 0.15 ($T_c$ = 41 K, $T$ = 23 K), $x$ = 0.12 ($T_c$ = 29 K, $T$ = 12 K), and $x$ = 0.10 ($T_c$ = 27 K, $T$ = 16 K).  The shaded area marks the energy (0.5 eV) scale needed to verify the FGT sum rule. In the right panel $(\pi/2)\omega \sigma_2^{ab}$ is approximately independent of $\omega$ for the three samples, as expected, and for $\omega \to 0$ provides the spectral weight of the condensate.} 
\label{Delta}
\end{figure}

The behavior of the spectral weight in the superconducting phase has been studied in the sample with $x$ = 0.12 and, for comparison, in other two LSCO single crystals with $x$ = 0.10 and 0.15. Their raw data are reported in Ref. \onlinecite{Venturini} and Ref. \onlinecite{Lucarelli}, respectively.  We plot in Fig.\ \ref{Delta} the difference $W_n  - W_s $ between the spectral weight in the normal phase at 50K and that at a $T$ well below $T_c$. The right-hand panel shows plots of $(\pi/2) \omega \sigma_2^{ab} (\omega)$, where  $\sigma_2^{ab}$ is the imaginary part of the in-plane conductivity and is approximately constant, as expected, for $\omega \leq$ 250 cm$^{-1}$. Its limit for $\omega \to 0$ gives the spectral weight which, below $T_c$, condenses at $\omega$ = 0.\cite{Molegraaf,Santander02} In the Figure, it coincides for the three samples with the difference $W_n - W_s$ at $\Omega \simeq 4000$ cm$^{-1}$ ($\sim$ 0.5 eV). Therefore in LSCO, as already reported for other hole-doped superconductors,\cite{Molegraaf} the energy range involved in the FGT sum rule is larger than expected for a conventional superconductor by one order of magnitude.

From $W_n - W_s$ one can also extract the London penetration depth $\lambda_L$. For the three crystals with $x$ = 0.10, 0.12, and 0.15 of Fig.\ \ref{B} one obtains 295, 280, and 215 nm, respectively. These values are in good agreement with a previous optical determination in LSCO with $x$ = 0.15\cite{Tanner93} and also with recent muon-spin-rotation measurements.\cite{Kadono} 

In conclusion, we have used model-independent quantities to compare the infrared response of an underdoped LSCO superconductor and a non-superconducting LSCO crystal (indicated in the phase diagram of LSCO as a "normal metal") with that of a conventional metal like gold. In all the three samples, the spectral weight $W (\Omega, T)$ follows the quadratic law $W = W_0 - B(\Omega)T^2$ for any $\Omega$ lower than $\omega_p$. This allows one to introduce a useful quantity, the "thermal response" $B(\Omega)$. In gold, $B(\omega_p)$ is reduced to a vanishingly small fraction of its low-$\omega$ value, while in both LSCO samples and in BSCCO $B(\omega_p)$ is much different from zero. This confirms that the behavior of cuprates is dominated by correlation effects, which extend the energy scale of the restricted sum rule much beyond the plasma edge. We have then applied a single-band hopping model to verify whether $W_0$ and $B$ are controlled by the same energy scale, as predicted for normal metals. The result clearly indicates the coexistence of two different energy scales in both LSCO samples, $t_0$ and $t_T$. $t_0$ is consistent with the bandwidth of photoemission experiments and scales with doping, while $t_T$ is smaller by one order of magnitude. As $t_T$ controls the transfer of spectral weight triggered by temperature and has the right size ($\sim$ 20 meV or 200 K), it could be related to phenomena like the opening of pseudogaps. We intend to further develop our analysis to understand if the marked difference observed in $B(\Omega)$ at low energy between the underdoped and the "normally metallic" LSCO may support this hint.
Concerning the superconducting phase, the spectral weight lost below $T_c$ at $\omega > 0$ is fully recovered, within errors, by the weight condensed at $\omega$ = 0 in a spectral range of about 0.5 eV. This value is lower than in cuprates with higher $T_c$'s, but higher by one order of magnitude than in conventional superconductors. The London penetration depth depends on doping and is consistent with muon-spin-rotation values.

We are indebted to L. Benfatto, M. Capone, C. Castellani, M. Grilli, J. Lorenzana, and A. Toschi for many useful discussions. We also wish to thank M. Fujita, K. Yamada, N. Kikugawa and T. Fujita for providing the LSCO crystals whose optical data have been reanalyzed in the present work.

\end{document}